\documentclass{jfm}
\usepackage{booktabs}  
\usepackage{graphicx}
\usepackage{subfig}
\usepackage{natbib}
\usepackage{lipsum}
\usepackage{hyperref}
\hypersetup{%
    colorlinks,
    citecolor=black,
    filecolor=black,
    linkcolor=black,
    urlcolor=black
}
\usepackage{amsmath}
\usepackage{caption}

\emergencystretch=\maxdimen
\ifCUPmtlplainloaded \else
  \checkfont{eurm10}
  \iffontfound
    \IfFileExists{upmath.sty}
      {\typeout{^^JFound AMS Euler Roman fonts on the system,
                   using the 'upmath' package.^^J}%
       \usepackage{upmath}}
      {\typeout{^^JFound AMS Euler Roman fonts on the system, but you
                   dont seem to have the}%
       \typeout{'upmath' package installed. JFM.cls can take advantage
                 of these fonts,^^Jif you use 'upmath' package.^^J}%
      }
  \else
  \fi
\fi

\ifCUPmtlplainloaded \else
  \checkfont{msam10}
  \iffontfound
    \IfFileExists{amssymb.sty}
      {\typeout{^^JFound AMS Symbol fonts on the system, using the
                'amssymb' package.^^J}%
       \usepackage{amssymb}%

      }{}
  \fi
\fi

\ifCUPmtlplainloaded \else
  \IfFileExists{amsbsy.sty}
    {\typeout{^^JFound the 'amsbsy' package on the system, using it.^^J}%
     \usepackage{amsbsy}}
    {\providecommand\boldsymbol[1]{\mbox{\boldmath $##1$}}}
\fi

\newcommand\Pran{\mbox{\textit{Pr}}} % Prandtl number, cf TeX's \Pr product
\newsavebox{\astrutbox}
\sbox{\astrutbox}{\rule[-5pt]{0pt}{20pt}}

\title[Mixed boundary conditions in Rayleigh-B\'enard convection]{Mixed
insulating and conducting thermal boundary conditions in Rayleigh-B\'enard
convection}

\author[D. Bakhuis, R. Ostilla-M\'onico, E. P. van der Poel, R. Verzicco and
D. Lohse]
{Dennis Bakhuis$^1$,\ns Rodolfo Ostilla-M\'onico$^{1,2}$,\ns
Erwin P. van der Poel$^1$,\break
Roberto Verzicco$^{1,3}$%
and Detlef Lohse$^1$}

\affiliation{%
$^1$Physics of Fluids Group, Department of Science and Technology, Mesa+
Institute, and J. M. Burgers Center for Fluid Dynamics, University of Twente,
7500 AE Enschede, The Netherlands\\
$^2$School of Engineering and Applied
Sciences and Kavli Institute for Bionano Science and Technology, Harvard
University, Cambridge, MA 02138, USA\\
$^3$Dipartimento di Ingegneria Industriale, University of Rome `Tor Vergata',
Via del Politecnico 1 Roma 00133, Italy}

\pubyear{2016}
\volume{650}
\pagerange{119--126}
\date{?; revised ?; accepted ?. - To be entered by editorial office}
\newcommand\Ray{\mbox{\textit{Ra}}}  % Rayleigh number
\newcommand\Nus{\mbox{\textit{Nu}}}  % Nusselt number

\usepackage{standalone}
\usepackage{
  booktabs, %\toprule, \midrule, \botrule
  amsmath, %\text
  siunitx, % units plotting
  pgfplotstable
}
\begin{document}

\maketitle

\begin{abstract}
A series of direct numerical simulations of Rayleigh-B\'enard convection, the
flow in a fluid layer heated from below and cooled from above, were conducted
to investigate the effect of mixed insulating and conducting boundary
conditions on convective flows.  Rayleigh numbers between $\Ray=10^7$ and
$\Ray=10^9$ were considered, for Prandtl numbers $\Pran=1$ and $\Pran=10$.
The bottom plate was divided into patterns of conducting and insulating
stripes.  The size ratio between these stripes was fixed to unity and the
total number of stripes was varied.  Global quantities such as the heat
transport and average bulk temperature and local quantities such as the
temperature just below the insulating boundary wall were investigated.  For
the case with the top boundary divided into two halves, one conducting and one
insulating, the heat transfer was found to be approximately two thirds of the
fully conducting case.  Increasing the pattern frequency increased the heat
transfer which asymptotically approached the fully conducting case, even if
only half of the surface is conducting.  Fourier analysis of the temperature
field revealed that the imprinted pattern of the plates is diffused in the
thermal boundary layers, and cannot be detected in the bulk.  With
conducting-insulating patterns on both plates, the trends previously described
were similar, however, the half-and-half division led to a heat transfer of
about a half of the fully conducting case instead of two-thirds.  The effect
of the ratio of conducting and insulating areas was also analyzed, and it was
found that even for systems with a top plate with only $25\%$ conducting
surface, heat-transport of $60\%$ of the fully conducting case can be seen.
Changing the 1D stripe pattern to 2D checkerboard tessellations does not
result in a significantly different response of the system.
\end{abstract}


\section{Introduction} Natural convection is a common and important phenomenon
which is omnipresent in Nature.  It leads to the transfer of internal energy,
within an unstably stratified fluid layer, via a buoyancy induced flow.  Ocean
currents, which are driven by gradients in density and salinity
\citep{mar99,wir06}, and the mantle convection inside the Earth, which drives
the plate tectonics and generates the geomagnetic field \citep{mck74,gla95},
are two examples of natural convection.  Even outside of our planet, at the
most distant stars, convection is of tremendous importance
\citep{spi71,cat03}.

An idealized system that is commonly used to study natural convection, as it
is mathematically well-defined and can be reproduced in a laboratory
experiment, is  Rayleigh-B\'enard (RB) convection \citep{nor77, ahl09, loh10,
chi12}.  The RB system consists of a fluid in a container, that is heated from
below and cooled from above.  The fluid is subject to an external
gravitational field $g$.  Apart from the geometric ones, this system has two
non-dimensional control parameters, namely the Rayleigh number $\Ray = \beta g
\Delta H^3 / \nu \kappa$, which measures the strength of the thermal driving,
and the Prandtl number $\Pran = \nu / \kappa$, a property of the fluid, where
$\beta$ and $\kappa$ are the isobaric thermal expansion and temperature
diffusivity coefficients of the fluid, $H$ the system height, $\Delta$ the
applied temperature difference between the plates, and $\nu$ the kinematic
viscosity.  Depending on the geometry of the system, other control parameters
such as the aspect ratio of the system, $\Gamma = L / H$ appear, where $L$ is
a characteristic horizontal length of the system.

Above a certain critical Rayleigh number, RB flow is linearly unstable, and
any perturbation will cause the onset of convection. This critical value is
determined by the properties of the fluid and the boundary conditions (BC) of
the RB system. If the thermal driving of the system is far
above the critical $\Ray$, the flow becomes turbulent. This dramatically
increases the heat transfer with respect to the purely conductive case.
Modeling this heat transfer is essential for understanding what is going on
inside stars, the Earth's mantle and many other systems. RB experiments (and
simulations) typically consist of a bottom and a top plate which have
homogeneous boundary conditions, and lateral boundary conditions which are
either periodic (simulations) to mimic laterally unconfined systems or
adiabatic (experiments) to account for a lateral confinement that minimizes
the heat losses.

However, these idealized systems assume that both the top and bottom plates
have perfectly homogeneous conducting surfaces while for all real physical
systems, there is a certain degree of imperfection.  In Nature we see such
inhomogeneities; for example, the fractures in ice floes \citep{mar12} or the
much debated insulating effects of continents on mantle convection
\citep{len05}.  Other examples include convection over mixed (agricultural)
vegetation and cities \citep{zhao14}.  In engineering applications, or in RB
experiments these can be small defects or dirt at the conducting plates, which
could result in lower than optimal heat transport. The limiting
cases of the boundary conditions are constant heat flux boundary
conditions, constant temperature boundary conditions, or thermally
insulating boundary conditions with no heat flux.  The difference in heat
transfer between the first two types of boundary conditions, Dirichlet and
von Neumann, was found to be negligible at large $\Ray$ DNS \citep{joh07,
joh09, ste11}, but the increase in flow strength under fixed-temperature
BC cannot be neglected \citep{huang17}. Accounting for a finite conductivity
of the thermal sources can lead to significant reduction in the heat transport
\citep{ver04}.

Temperature boundary conditions can also be spatially and temporally varying.
The study of the effects of these imperfect boundary conditions goes back to
\cite{kel78} and later \cite{yoo91} who applied sinusoidal temperature
boundary conditions on both plates, which mimic plates with embedded heaters.
These plates are locally hotter, when closer to the heater elements.
\cite{jin08} showed that the Nusselt number is increased when the energy input
into the system from the plates is periodically pulsed instead of stationary.
Recent experiments from \cite{wang17} using insulating lids at the top
boundary of a RB cell showed that with increasing insulating fraction the same
amount of heat goes through a smaller cooling area.   Some
simulations of inhomogeneous boundary conditions have also been
performed. \cite{cooper13} simulated two and three-dimensional
Rayleigh-B\'enard systems of mixed adiabatic-conducting boundary
conditions at one plate at moderate Rayleigh numbers, with a
geophysical focus, finding that the distribution of the patches caused
changes in the flow configuration, the bulk temperature and the
Nusselt number. The simulations by \cite{rip14} added non-conducting
defects in the form of periodic patches to the top plate of a
two-dimensional numerical RB system, and studied both the transition
to turbulence of RB flow, finding a delay in this transition when
defects were present, and the fully turbulent regime, finding a
decrease in Nusselt number when the patch wavelength was larger than the
characteristic thermal boundary layer scale. 

Here, we extend the research of \cite{rip14} by applying non-conducting stripe
patterns to a three-dimensional RB system.  We will focus on the fully
turbulent regime instead of the transition to turbulence,
and consider a wider range of patterns at higher $\Ray$, extending 
the work by \cite{cooper13}.  We start by
applying distributions of striped insulating patterns to the top boundary
only, and study the dependence of both local and global variables, e.g.
effective heat transfer and average bulk temperature, on the number of
stripes.  For most of the study, we keep the conducting to insulating areas
equal to each other, but the effect of this ratio is also studied, which
mimics the degree of imperfections and pollution on the plates.  We also study
the effect of applying the same pattern to both plates and the role of the
pattern geometry by applying a two-dimensional checkerboard
insulating-conducting pattern on the top plate.  Checkerboards and stripes can
be seen as the two limiting cases for the pattern geometry.

This manuscript is organized as follows: First, in \autoref{NumericalMethod}
we detail the geometry and numerical method.  In the next section, the results
for the stripe pattern variations will be discussed where both conducting and
insulating areas are kept constant.  This is first done on the top plate and
later the pattern is applied on both plates.  A Fourier analysis was performed
to study the penetration depth of the pattern imprint in the flow.  In
\autoref{BothPlates} a pattern is added to both plates, in
\autoref{VariationFraction}, we present and discuss the results for varying
the ratio of conducting to insulating surface while keeping the number of
divisions constant and in \autoref{mixedTwoDimensions} we present the results
for a plate with a checkerboard pattern instead of a striped pattern.  The
manuscript is concluded by presenting the conclusions and outlook in
\autoref{Conclusion}.

%%%%%%%%%%%%%%%%%%%%%%%%%%%%%%%%%%%%%%%%%%%%%%%%%%%%%%%%
\section{Numerical method}\label{NumericalMethod}
%%%%%%%%%%%%%%%%%%%%%%%%%%%%%%%%%%%%%%%%%%%%%%%%%%%%%%%%

In this numerical study, we solve the incompressible Navier-Stokes equations
within the Boussinesq approximation for RB. In non-dimensional form, these
read:
\begin{equation} \begin{aligned} \frac{D \boldsymbol{u}}{Dt} &= -\nabla P +
    \left( \frac{\Pran}{\Ray} \right)^{1/2} \nabla^2 \boldsymbol{u} + \theta
    \hat{z}, \\ \frac{D \theta}{Dt} &= \frac{1}{\left( \Pran \Ray
    \right)^{1/2}} \nabla^2 \theta, \\ \nabla \cdot \boldsymbol{u} &= 0,
\end{aligned} \end{equation}
where $u$ is the non-dimensional velocity, $P$ is the non-dimensional
pressure, $\theta$ is the non-dimensional temperature, and $\hat{z}$ is the
unit vector pointing in the direction opposite to gravity $g$.  For
non-dimensionalization, the temperature scale is the temperature difference
between the plates $\Delta$, the length scale their distance $H$ and the
velocity scale is the free-fall velocity $U_f = \sqrt{g\beta \Delta H}$.

We consider a geometry which is a horizontally doubly-periodic cuboid. The
domain has horizontal periodicity lengths of $L_x$ and $L_y$, and a vertical
dimension $H$.  These variables with a tilde superscript denote their
non-dimensional counterparts.  The equations were discretized using an
energy-conserving second-order finite-difference scheme, and a fractional
time-step for time marching using a third-order low-storage Runge-Kutta scheme
for the non-linear terms, and a second order Adams-Bashworth scheme for all
viscous and conducting terms \citep{ver96,poe15}.  The code was heavily
parallelized to run on hundreds or even thousands of cores simultaneously and
was validated many times \citep{ste10,ste11, poe15}.  Recently, the code was
open-sourced and is available for download at \url{www.AFiD.eu}.

The domain was discretized by $n_x \times n_y \times n_z = 360 \times 360
\times 288$ grid points.  In both horizontal directions, the grid was
uniformly divided, and in the vertical direction the points were clustered
near the top and bottom plates.  A number of simulations were conducted to
test the aspect ratio dependence and the grid independence.  These tests did
not show any significant differences in the range of Rayleigh numbers used in
this study.  For RB convection, a series of exact relationships which link the
Nusselt number to the global kinetic energy dissipation ($\nu \nabla^2 u$) and
the thermal dissipation ($\kappa \nabla^2 \theta$) exist \citep{shr90,ahl09},
and they have been further used to check the spatial accuracy of the
simulation as in \cite{ste10}.  The size of the time steps was chosen
dynamically by imposing that the Courant-Friedrichs-Lewy (CFL) number in the
grid would not exceed $1.2$.%

The main response of the system is the Nusselt number ($\Nus$), which is the
heat transfer non-dimensionalized using the purely conductive heat transfer:%
\begin{equation}
    \Nus = \frac{\left< u_z \theta\right>_A - \kappa\partial_z
    \left<\theta\right>_A}{\kappa \Delta L_z}, 
\end{equation}
where $\left<\cdot \right>_A$ indicates the average over any horizontal plane.
The simulations were run between 50 and 100 large-eddy turnover times based on
$U_f$ and $H$. Statistical convergence is assessed by calculating
differences in $\Nus$ between final and half the amount of measurement
points. These are shown as error bars in the plots.

\begin{figure}
\centering
\includegraphics[scale=1]{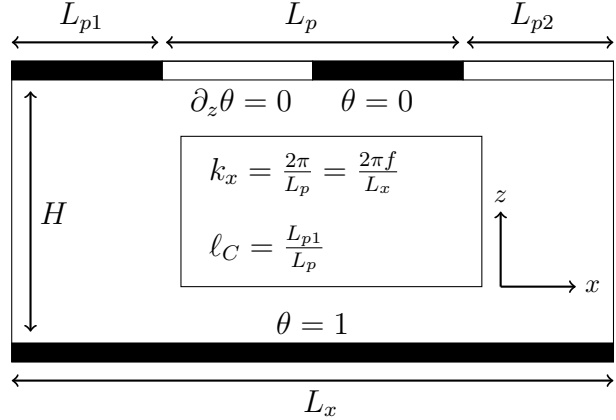}
\caption{%
Two-dimensional $y$-cut of the geometry. The domain has the
dimensions $L_x \times L_y \times H$. The bottom plate, at $\hat{z}=0$,
has $\theta=1$. The top plate is divided into stripes of conducting
($\theta=0$) and insulating ($\partial_z \theta = 0$) regions.}
\label{figure1}
\end{figure}

In the classical RB case, both the top and bottom plates have a homogeneous
boundary condition, i.e. they are perfectly conducting. Here, we use this
boundary condition only for the bottom plate, while top plate is taken to have
periodic patches of insulating regions which do not contribute to the heat
transfer from fluid to plate. The definition of these patches are similar to
those in \cite{rip14}:
\begin{equation} \begin{aligned} \theta(\hat{x},\hat{y},\hat{z}=1) &= 0 &&
    \forall \hat{x},\hat{y} \in \left[ j L_p,L_{p1}+j L_p \right], j \in
    \mathbb{Z} \\ \partial_z \theta(\hat{x},\hat{y},\hat{z}=1)  &= 0 &&
    \forall x,y \notin \left[ j L_p,L_{p2}+j L_p \right], j \in \mathbb{Z} \\
    \theta(\hat{x},\hat{y},\hat{z}=0) &= 1 && \forall x,y.  \end{aligned}
\label{eq:boundaryConditions} \end{equation}
Here $L_p$ is the width of a pair of patches.  $L_{p1}$ is the width of the
conducting part, $L_{p2}$ is the insulating part, and the hat on the spatial
coordinates indicates non-dimensionalizations.  For most of this study we keep
the insulating and conducting areas equal, i.e. $\ell_C=1/2$.  The BC on the
top plate depends only on the $x$-coordinate, which results in sets of
insulating and conducting stripes.  The number of stripe pairs in a horizontal
direction L were defined as $f = L / L_p$, which is a central control
parameter of this study.
A two-dimensional schematic is shown in Figure \ref{figure1}.

There are limitations on the value of $f$. As each stripe has to be an integer
number of grid points, only integer multiples of the grid resolution are
valid. In addition, the width of all stripes summed should fit inside the
system, e.g. the width in grid points should exactly be the number of grid
points in the x-direction. This results in a total of 18 different pattern
frequencies. For the smallest possible frequency, $f=1$, we have one
conducting and one insulating stripe, which both have a width of 180 grid
points. The pattern with the largest frequency, $f=90$, has 90 stripe pairs
where each stripe has a width of two grid points. In this paper we will use
the wavenumber $k_x = \left( 2 \pi f \right) / L_x$, to describe the stripe
distribution.

\begin{figure}
\centering
\includegraphics[width=0.46\textwidth]{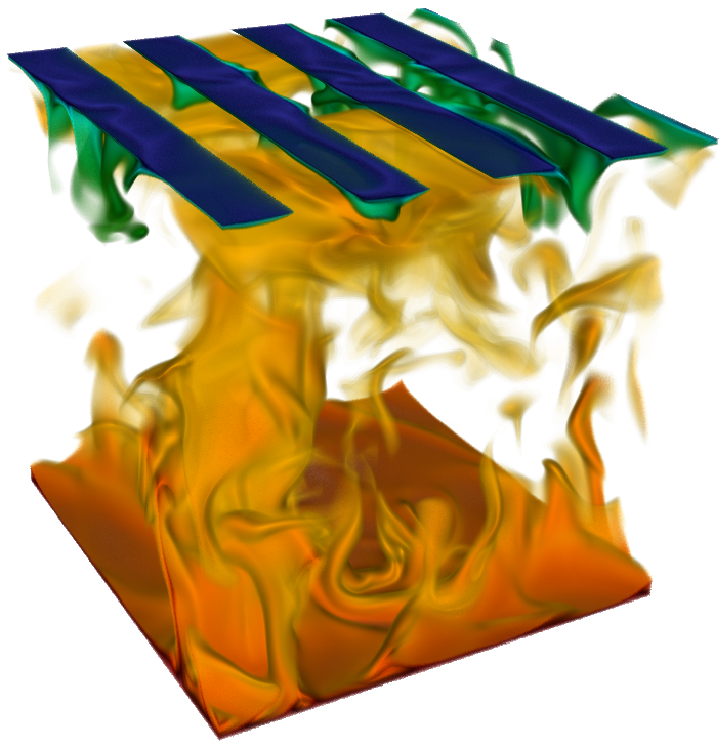}
\qquad
\includegraphics[width=0.46\textwidth]{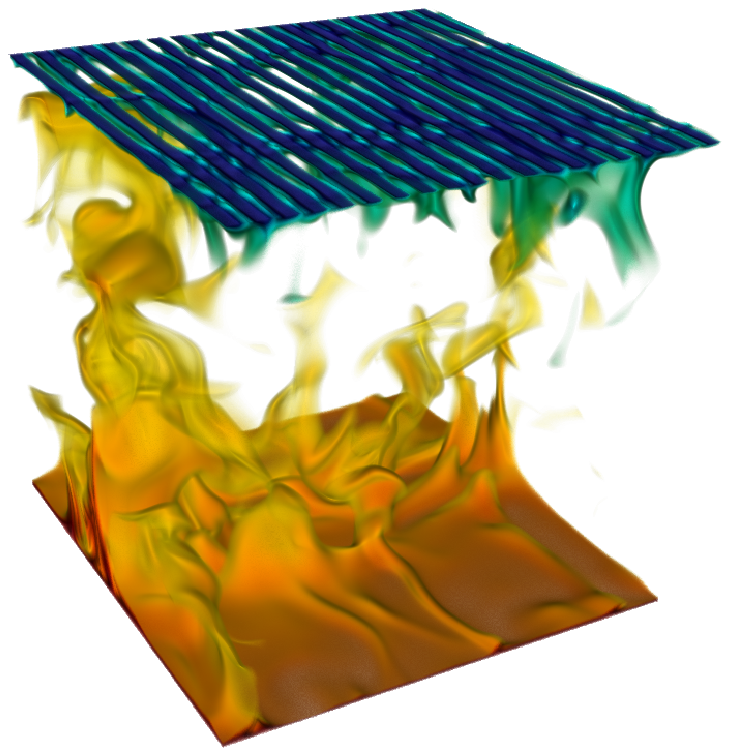}
\caption{%
Two 3D visualizations of the instantaneous temperature field with
different pattern frequencies applied to the top boundary.  $\ell_C=0.5$,
$\Ray=10^8$ and $\Pran=1$ for both cases.  Hot fluid is shown in red while
the cold fluid has a blue color.  The left visualization shows $f=4$, four
insulating stripes and four conducting stripes. The right visualization
shows five times as many stripe pairs, with $f=20$.  Plumes of colder
fluid are ejected primarily from the conducting areas for $f=4$ while for
$f=20$ the plumes also eject on areas below insulating regions.  }
\label{figure2}
\end{figure}

To give an idea of how the flow in such a system looks like, two different
instantaneous snapshots of the RB system with two different stripe frequencies
are shown in Figure \ref{figure2}. Both cases have exactly the same conducting
and insulating areas, but a different stripe pattern. In section
\ref{mixedTwoDimensions}, we also vary the BC in y-direction, while keeping
both the insulating and conducting areas equal, which results in a
checkerboard pattern.

%%%%%%%%%%%%%%%%%%%%%%%%%%%%%%%%%%%%%%%%%%%%%%%%%%%%%%%%%
\section{Results}\label{VariationPattern}%
%%%%%%%%%%%%%%%%%%%%%%%%%%%%%%%%%%%%%%%%%%%%%%%%%%%%%%%%%

\subsection{The effect of the number of stripes}\label{NumberofStripes}%

In this subsection we present results of a series of simulations in which we
varied the number of stripes while keeping $\ell_C = 1/2$. Four sets of cases
were run for $\Ray = 10^7$, $10^8$, and $10^9$ and $\Pran = 1$ and $10$.

%
%%%%%%%%%%%% Nus(Ra) %%%%%%%%%%%%%
%
\begin{figure}
\centering
\subfloat{%
\includegraphics{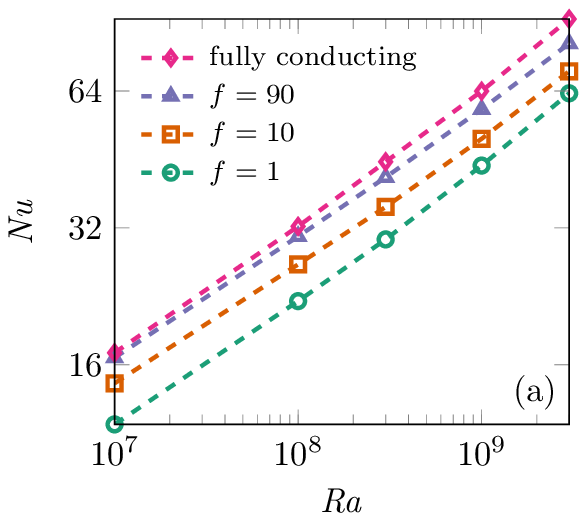}%
\label{figure:raVSnus}%
}
\subfloat{%
\includegraphics{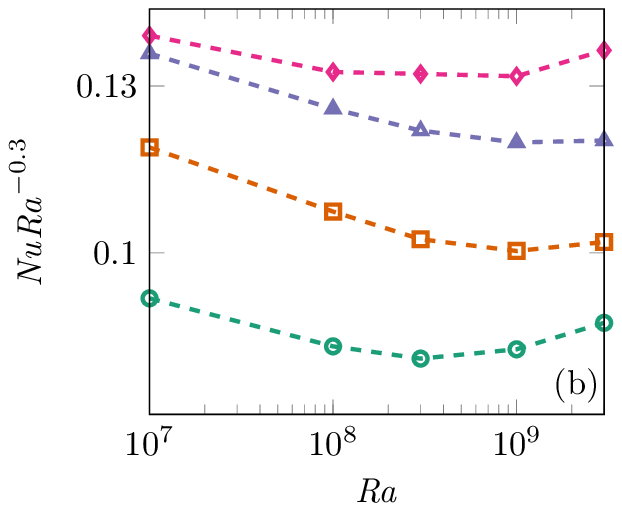}%
\label{figure:raVSnusra}%
}
\caption{%
(a) Nusselt number $Nu$  and (b)
compensated Nusselt number $NuRa^{-0.3}$  against $Ra$ for various $f$ and
$Pr=1$. The markers show the actual results while the dotted lines
indicate the trend between these simulations. $\ell_C$ was fixed to
$\ell_C=0.5$ for all simulations.}
\label{figure:raVSnusboth}%
\end{figure}%

First we show in Figure \ref{figure:raVSnus} the curves of $\Nus$($\Ray$) for
the fully conducting case and three striped cases with $f=$ 1, 10, and 90.
The markers show the actual results from the simulations and the dotted lines
indicate the trend between the measurements.  From this figure we see that the
scaling does not differ significantly between all cases and all curves are
almost parallel.  We do see a strong dependence on the wavenumber of the
pattern.  For $f=1$, we computed that $\Nus$ is approximately $2/3$ of the
fully conducting case.  Increasing $f$ results in a larger $\Nus$ and the
values almost converge with the fully conducting case.  Both the $f=90$ and
$f=10$ cases are relatively closer to the fully conducting case at $\Ray=10^7$
than at $\Ray=3 \times 10^9$. Our interpretation is that, due to the lower
$\Ray$, the thermal BL is thicker which results in larger horizontal
conduction of heat.  As all curves have approximately the same scaling we have
compensated the data using $\Ray^{-0.3}$, which is shown in figure
\ref{figure:raVSnusra}.  Here it is clearly visible that for low $\Ray$, the
$f=90$ case is almost as efficient as the fully conducting case, but the
differences increase with increasing $\Ray$.  The two extreme cases, $f=1$ and
the fully conducting case, follow a similar trend, however, the $f=1$ case is
shifted to a lower, less efficient level.  Both figures clearly show that an
increase in wavenumber of the pattern results in an increase in $\Nus$.

\begin{figure}
\centering
\subfloat{
\includegraphics{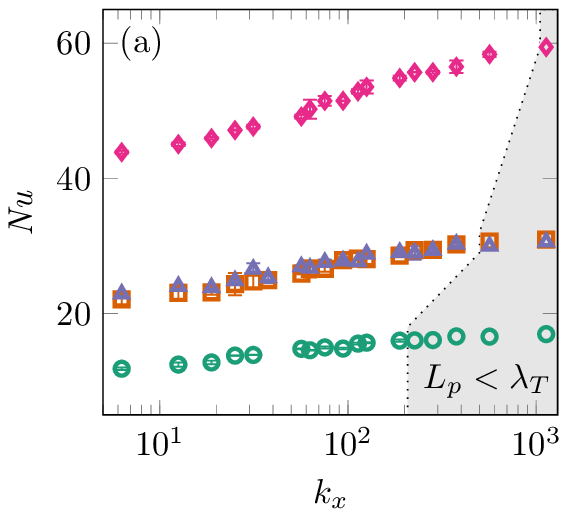}
\label{figure3a}
}
\subfloat{
\includegraphics{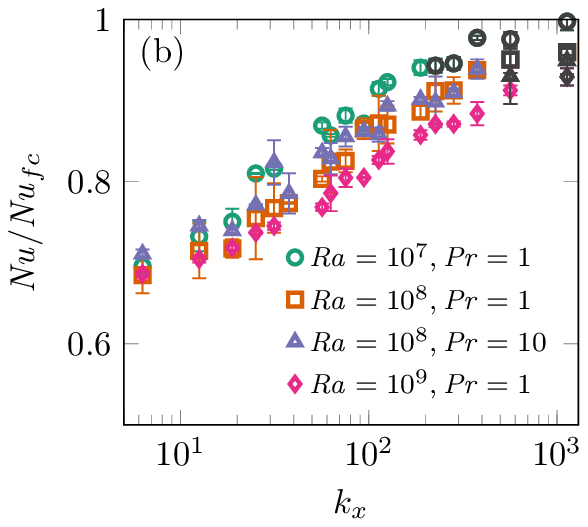}
\label{figure3b} }
\caption{%
(a) Nusselt number $\Nus$ for various $\Ray$
and $\Pran$ versus pattern wavenumber $k_x$. The shaded area shows the
data points for which the stripe width is smaller than the thermal
boundary layer. (b) The same data but normalized by $\Nus_{fc}$. The dark
points are the same as in the shaded area of 4a. For both plots, the error
bars shows the statistical convergence error.}
\label{figure3}
\end{figure}

To make this point even more clear, we plotted $\Nus$($k_x$) in figure
\ref{figure3a}.  All datasets show a clear $k_x$ dependence: $\Nus$ quite
strongly increases with $k_x$.  The gray area indicates the data points for
which the width of the stripes $L_p$ is smaller than the thermal boundary
layer thickness $\lambda_T$, which can be estimated as $\lambda_T = H / (2
\Nus)$.  \citeauthor{rip14} showed that the heat transfer monotonically
increases with increasing $k_x$ until $L_p$ is comparable to $\lambda_T$.  A
similar trend is visible for $\Ray=\num{1e7}$ and $\Ray=\num{1e8}$ even if the
$\Ray=\num{1e7}$ case shows minor increases in heat transfer when $L_p$ is
further decreased beyond $\lambda_T$.  The change in $\Pran$ from $\Pran=1$ to
$\Pran=10$ does not have a significant effect for the heat transfer, at least
not for $\Ray = 10^8$, for which the two datasets are practically overlapping.
This behavior is similar to the standard Rayleigh-B\'enard case, in which the
$\Pran$ of $\Nus$ is also weak \citep{ahl09}.

To perform a better comparison between the different $\Ray$, we normalize the
resulting $\Nus$ using $\Nus_{fc}$, the Nusselt number of the fully conducting
case.  The normalized $\Nus$ for different stripe configurations are shown in
figure \ref{figure3b}.  We see the same trend for all $\Ray$ and $\Pran$.  At
the lowest $k_x$, in which we only have a single conducting and single
insulating stripe, the effective $\Nus$ is approximately two-thirds of the
fully conducting case.  When the number of stripes, i.e. $k_x$, is increased,
we see that for all tested $\Ray$ and $\Pran$, the Nusselt number slowly
converges to almost the fully conducting case.  So remarkably even if only
half of the plate is conducting, it can be almost as effective as if the plate
is fully conducting.  In this compensated plot it is also clearly visible that
for the largest $k_x$ of $\Ray=\num{1e7}$, for which $L_p$ goes below the size
of $\lambda_T$, the heat transfer is still increasing.

%
%%%%%%%%%%%% Nus(kx) %%%%%%%%%%%%%
%

The differences between the different $\Ray$ are not so clear at the lowest
$k_x$; however, for a slight increase of $k_x$ we see that the curves order
themselves.  $\Ray = 10^9$ increases slightly slower with $k_s$ when comparing
it to the $\Ray = 10^8$ case.  The $\Ray = 10^7$ case increases slightly
faster than the $\Ray = 10^8$ case and it ends at 99.7\% of the fully
conducting case for the largest number of stripes used.  The explanation for
this trend is the difference in boundary layer thickness, which decreases when
we increase $\Ray$.  The boundary layer controls the heat transport from the
bulk to the conducting region.  Below the insulating region, this heat
transport must also go in the horizontal direction.  By increasing $\Ray$ and
thus decreasing the thickness of the boundary layer, the same amount of heat
needs to be conducted through a smaller 'channel'.  This explains the
decreased effectiveness with increasing $\Ray$.

In these sets of simulations, we also compare $\Pran = 10$ with $\Pran = 1$
for $\Ray = 10^8$.  The results for both $\Pran$ are quite similar.  At lower
$k_x$, we see that the $\Pran =10$ case has only a marginally larger $\Nus$
and these differences become smaller with increasing $k_x$ and are within the
uncertainty of the simulation.

\begin{figure}
\centering
\subfloat{
\includegraphics[width=0.48\textwidth]{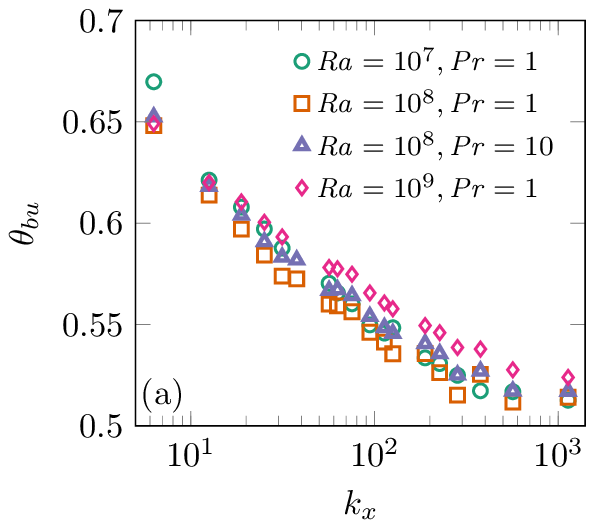}%
\label{figure4a}
}
%\quad
\subfloat{
\includegraphics[width=0.48\textwidth]{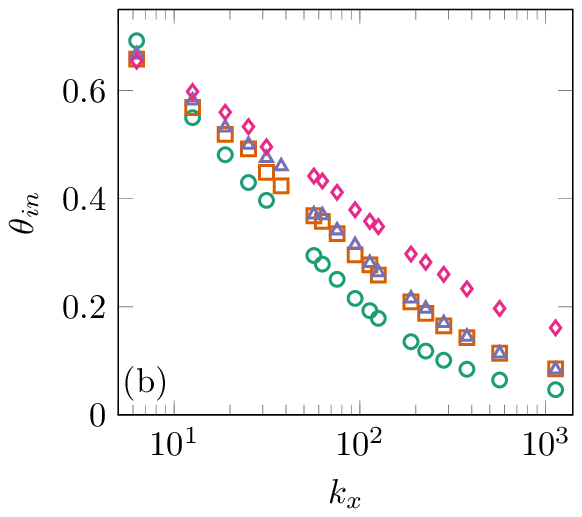}
\label{figure4b}
}
\caption{(a) Temperature of the bulk fluid $\theta_{bu}$
for various $\Nus$ and $\Pran$ versus pattern wavenumber $k_x$. (b)
Temperature below the insulating stripes $\theta_{in}$, averaged over the
entire insulating area, for different $\Ray$ and $\Pran$ against pattern
wavenumber $k_x$.}
\label{figure4}
\end{figure}

%
%%%%%%%%%%%% bulk temp / insulating region temp %%%%%%%%%%%%%
%
Figure \ref{figure4a} shows the average bulk temperature of the fluid plotted
against the wavenumber.  This average was computed over the horizontal plane
at mid-height of the system.  The effect is similar to what we see for
$\Nus/\Nus_{fc}$ in figure \ref{figure3b}.  At the lowest $k_x$ the average
bulk temperature has increased to about $2/3$.  In that case top plate is
split in half and only one half is contributing to heat transfer.  While
ignoring the adiabatic area, we can divide the conducting areas into three
equally sized parts with two parts on the bottom plate and one on the top
plate.  Using the same reasoning to that used for the symmetric case, we
obtain the following equality for the average bulk temperature: $\theta_{bulk}
= \frac{2}{3} \theta_{bottom} + \frac{1}{3} \theta_{top} = 2/3$.  As with
$\Nus/\Nus_{fc}$, we see that the average bulk temperature approaches the
fully conducting case of $\theta_{bulk}=1/2$ for increasing $k_x$.  When
comparing the various curves, they all appear similar, with a maximum
difference of about 0.02 to 0.03 in the average bulk temperature.
Unfortunately, \cite{wang17} do not report their average
temperature and a comparison to their calculations with large-wavelength
imperfect boundary conditions is impossible.

In figure \ref{figure4b} we show the horizontally and time averaged
temperature below the insulating area of the top plate as function of the
wavenumber.  At the lowest wavenumber (top plate split into equal conducting
and insulating regions), we see that the averaged temperature is almost equal
for all $\Ray$ and $\Pran$.  When increasing $k_x$, a dependence on $\Ray$
emerges.  After just a few additional divisions in stripes, all curves order
themselves according to $\Ray$, with the lower $\Ray$ values approaching the
lower bounds faster than the larger ones.  At the largest $k_x$, the
temperature difference between $\Ray =10^9$ and $\Ray=10^7$ is 15\%.  As the
temperature below the insulating area is slightly higher than for the area
below the conducting area because of lack of cooling, we can conclude that for
larger $\Ray$, the whole top layer is, on average, hotter than for the lower
$\Ray$.

\begin{figure}
\centering
\includegraphics[width=0.46\textwidth]{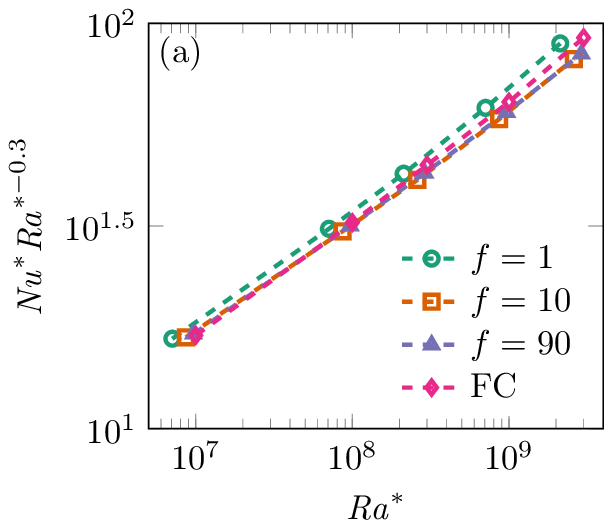}
\includegraphics[width=0.46\textwidth]{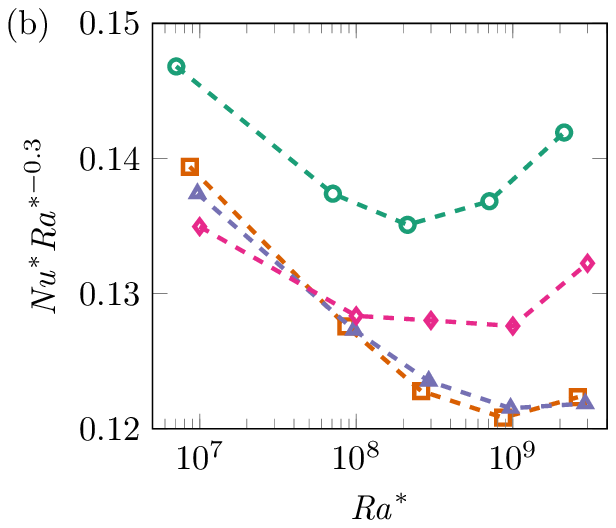}
\caption{%
(a) Corrected Nusselt number against compensated Rayleigh number
for selected pattern wavelengths. (b): Compensated and corrected
Nusselt number against corrected Rayleigh number.
No collapse, or natural ordering of the 
curves can be seen.  }
\label{figure5}
\end{figure}

These results suggest that it could be possible to account for the $\Nus(k_x)$
relationship by using corrected non-dimensional variables in the spirit of
\cite{ver04}. However, changing the (effective) thermal conductivity, ignores
the heterogeneities of the plate which is the main source of the observed
behavior. In figure \ref{figure5}, we show the corrected Nusselt number
$\Nus^*=\Nus/\theta^*$ against the corrected Rayleigh number $\Ray^*=\Ray
\theta^*$, where $\theta^*$ is the non-dimensional average temperature
difference given by $\theta^*=L_{p2}\theta_{in}/L_p$. No logical ordering can
be seen, and as expected, an attempt to describe the results with some global
effective thermal conductivity fails.

%
%%%%%%%%%% 2d fft
For horizontal slices of the instantaneous temperature field, a discrete
two-dimensional Fourier transform can be applied, which is defined as:
\begin{equation} \Theta(\gamma) = \left| \left< \sum\limits_{y=0}^{n_y-1}
\sum\limits_{x=0}^{n_x-1} e^{-i\left[ (2\pi / n_x)j_x x + (2\pi/n_y)j_y
y\right]} \theta(x,y,t) \right>_{t} \right|_{j_x=\gamma,j_y=0} \end{equation}
where $\left< . \right>_t$ is the time average and the $j_y$ mode was set to
zero, leaving a single wavenumber parameter $\gamma \equiv j_x$.
Using the described method on the horizontal slice just below the top boundary
we can identify the imprint the stripe structured BC leave on the flow.

\begin{figure}
\centering
\includegraphics{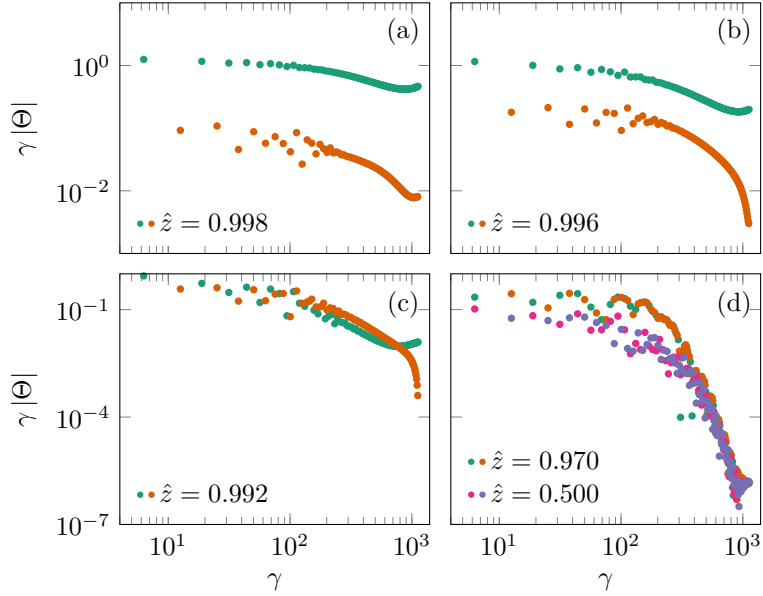}
\caption{%
Premultiplied two-dimensional Fourier transform of a horizontal slice at
various distances from the top boundary wall, averaged in time.  Except
for figure \ref{figure7}d, each figure is compiled using a single dataset.
The colors help to identify the different modes which are present in each
single dataset and are identified by the odd or even value.  a) At the closest
gridpoint ($\hat{z}=0.998$), the two different modes are clearly visible.  b)
One gridpoint further ($\hat{z}=0.996$), the distinction slowly fades.  c)
While still inside the boundary layer at $\hat{z}=0.992$, both modes are
practically overlapping.  d) Just outside the boundary layer, it is impossible
to distinguish two different modes at all.  As a reference, we also show the
spectrum at the centre of the system ($\hat{z}=0.5$).  }
\label{figure7}
\end{figure}

Using the Fourier transform we one identify the imprint of the BC just below
the top boundary in the flow itself.  Using this distinct signature we can
find out how far this pattern is still present once one moves away from the
boundary wall.  The distance from the top boundary wall is indicated using
$\hat{z}$.  In figure \ref{figure7}, we see the compensated spectra for $f=1$
at five different planes for increasing $\hat{z}$.  The colors are used to
identify the different modes and except for figure \ref{figure7}d)
are compiled using a single dataset (odd or even value). 
When moving away from the wall, we see that the two distinct
modes approach each other and just outside the boundary layer, at
$\hat{z}=0.970$, it is hard to distinct the two different curves at all.
Within the boundary layer the signature of the pattern almost completely fades
away and in the bulk flow is not visible at all.  The difference between the
Fourier transform just outside the BL ($\hat{z}=0.970$) and at mid-height of
the system ($\hat{z}=0.5$) is marginal.  These findings hold for the complete
range of $f$.

It is quite remarkable that even for the most extreme case at $f=1$, the
pattern is not visible in the bulk region.  This means that in the boundary
layer, in which conduction dominates, the temperature differences of the top
plate are averaged such that an effective, slightly higher cold plate is seen
by the bulk flow.

%
%%%%%%%%%% double-sided pattern
\subsection{Patterns on both plates}\label{BothPlates}
Until now we only applied the insulating and conducting patches to the top
boundary.  This resulted in a normalized heat transfer of approximately
two-thirds for the lowest $k_x$ and almost the fully conducting case at the
highest $k_x$.  Using a simple argument we could indeed rationalize the value
of two-thirds for the normalized heat transfer for the lowest wavenumber.  If
we now apply the same pattern also on the bottom plate, can we still get to
the same efficiency as if we only applied the pattern to the top boundary
wall?

\begin{figure}
\centering
\subfloat{%
\includegraphics{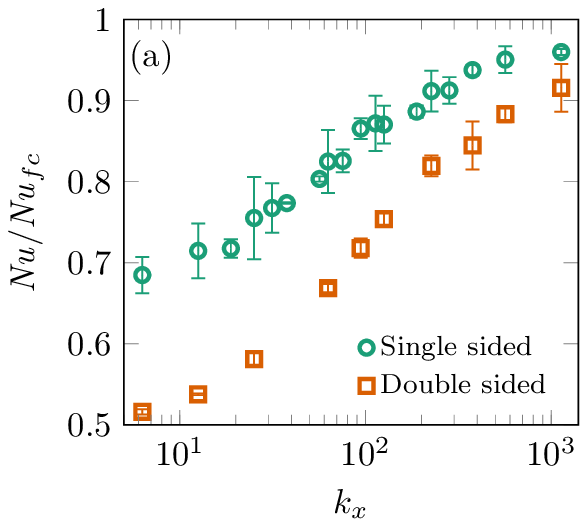}%
\label{figure8a}
}
%\qquad
\subfloat{%
\includegraphics{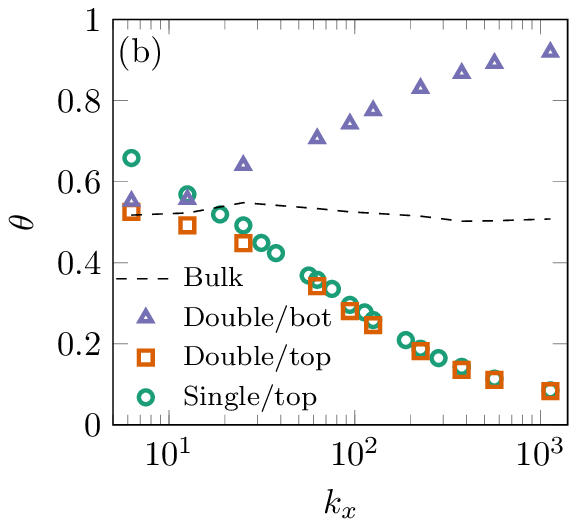}
\label{figure8b}
}
\label{figure8}
\caption{%
(a) Comparison of the normalized Nusselt
number $\Nus/\Nus_{fc}$ for the single and double-sided cases with
$\Ray=10^8$, $\Pran=1$, and $\ell_C=0.5$.  For the lowest $k_s$ we see
that the system is about $2/3$ and $1/2$ of the fully conducting case for
the single and double-sided case, respectively.  (b) Average temperature
above or below the insulating stripes for the single- and double-sided
case with $\Ray=10^8$, $\Pran=1$, and $\ell_C=0.5$.  The dashed line shows
the average bulk temperature, calculated at mid-height for the double-sided
system.}
\end{figure}

The comparison of the normalized Nusselt number $\Nus/\Nus_{fc}$ between the
single- and double-sided case is shown in figure \ref{figure8a}.  For the
lowest $k_x$ we see that the double-sided case conducts heat at approximately
half the rate of the fully conducting case.  This is in line with the
expectations, as we only have half the effective area on both boundary walls.
What also is visible is that the curve for the double-sided case is steeper
than the curve of the single-sided case and therefore reducing the difference
between both cases with increasing $k_x$.  For the largest $k_x$ the heat
transfer of this system again is almost as efficient as if it were fully
conducting.  There is only half of the area available for heat entering the
system and only half the area for heat leaving the system.  Still, the same
amount of heat transfer as if the system were fully conducting is achieved.
For the single-sided case as for the double-sided one, for the lowest $k_x$
the efficiency is not exactly $2/3$ and $1/2$ but slightly above these values.
The geometry of the double-sided case can be decomposed into a regular
Rayleigh-B\'enard cell and a neutral domain, both with identical dimensions,
positioned next to each other.  The top and bottom boundaries from this
neutral domain are both insulating and heat can only enter and exit from the
sides.  As we have periodic BC in the horizontal direction, both sides of the
regular RB area are connected to this neutral domain which acts as a buffer
for heat.  This extra buffer is the only difference and thus the cause for the
small difference.

The average temperature just below or above the insulating boundaries are
shown against $k_x$ in figure \ref{figure8b}.  The difference in temperature
below the top insulating boundary between the single-sided case, shown in
green cirlces , and the double-sided case shown in orange squares is only
significant at the lowest $k_x$.  Only at the lower $k_x$, the single-sided
system is hotter, just below the boundary.  At larger $k_x$, the asymmetry
does not make a difference on the temperatures and both temperatures converge
to the conducting plate temperature.  In the same plot, we also show the
temperature just above the insulating area of the bottom plate and the bulk
temperature.  For the double-sided case for the lowest $k_x$, the top, bottom,
and bulk temperatures are very similar.  This indicates that for the lowest
$k_x$ all the fluid in neutral domain, the large area confined by the
insulating bottom and top plate, has approximately the same temperature and
hardly contributes to the heat transfer.  This fully agrees with
$\Nus/\Nus_{fc}\approx 0.5$ seen in figure \ref{figure8a}.  The bulk
temperature of the double-sided case stays approximately 0.5 for the full
range of $k_x$, as must hold for a symmetric system.  Temperatures above the
bottom boundary and below the top boundary are also symmetric with respect to
the bulk temperature confirming statistical convergence of our calculations.
As for the single-sided case, figure \ref{figure8b} shows that for the largest
$k_x$ the temperature close to the insulating areas are very close to their
conducting counterpart.  The heat conduction in the boundary layer makes the
bulk fluid see an almost perfect heat conductor.

%
%%%%%%%%%% Varying the insulating fraction
\subsection{Variation of the insulating fraction}\label{VariationFraction} All
systems which we discussed until now had a conducting area with the size equal
to the insulating area, i.e. $\ell_C = 1/2$.  We can now vary $\ell_C$ and
look into its effect on the heat transfer.  In the previous simulations we
used $k_x$ as the wavenumber and this sets the number of insulating and
conducting stripes in the two-dimensional case.  The width of the system,
$L_x$, was divided in $f$ equal pairs of these stripes.  In the previous
subsection, these divisions were of equal areas.  Now we will change the ratio
of areas to make the top plate less and less conducting.  For these
simulations we fixed $k_x = 9$, $\Ray=10^8$ and $\Pran=1$.  Then the ratio
between the insulating and conducting area was varied, namely we simulated
$\ell_C = 1.0, 0.875, 0.75, 0.625, 0.50,$ and $0.25$, thus gradually reducing
the conducting area of the top plate from 85\% to 25\%.

\begin{figure}
\centering
\subfloat{%
\includegraphics{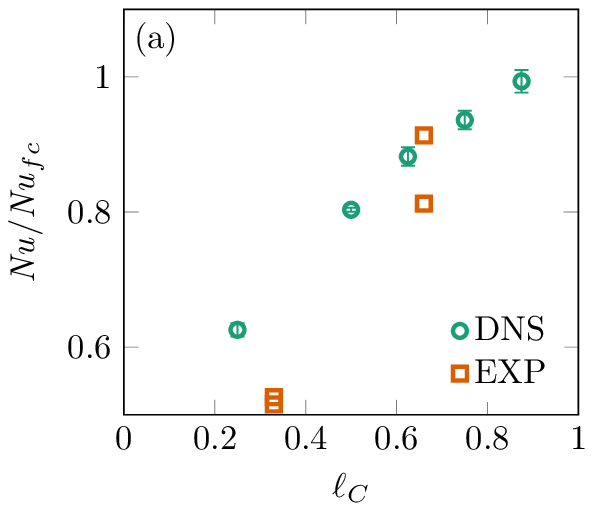}%
\label{figure9a}
}
\subfloat{
\includegraphics{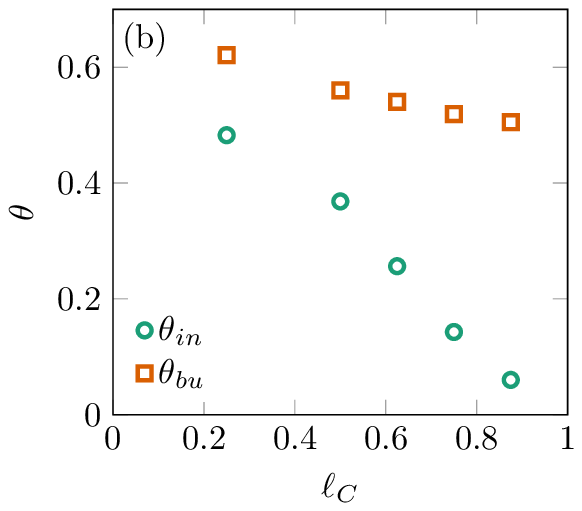}%
\label{figure9b}
}
\caption{%
(a) Normalized Nusselt number
$\Nus/\Nus_{fc}$ for various fraction $\ell_C$ of the conducting plate,
while keeping the pattern frequency $f=9$, $\Ray=10^8$, and $\Pran=1$ (green circles),
and the rectangular tank of \cite{wang17} (orange squares).
(b) Average temperature just below the insulating area $\theta_{in}$ and
average bulk temperature $\theta_{bu}$, both plotted against $\ell_C$.
Other parameters are the same as for (a).}
\label{figure9}%
\end{figure}

{ Figure \ref{figure9} shows the results of the simulations, as well 
as the data from the rectangular tank of from the experiments in \cite{wang17}
(extrapolated from Table 1).}
In the case where
only 15\% of the area is insulating we see that the difference with the fully
conducting case is almost negligible and within the statistical error.
Increasing the amount of insulating area to 25\%, the heat transfer is still
more than 90\% of the fully conducting case.  Even at 50\% conducting region
we still get the effectiveness of 80\%.  At the largest ratio of 75\%
insulating fraction we are still above 60\% of the heat transfer of the fully
conducting case.  In other words, the effective area of the top plate is only
a quarter of the fully conducting case but still we get a system that is only
40\% less effective. 

{ The rectangular tank of \cite{wang17} has patches 
with dimensions comparable to the system size. Two data points for the
two different
wavelengths are available for each $\ell_C$. 
The data point with a smaller wavelength (`ACA' and `CAC' patterns)
corresponds to the higher values of $Nu$. While
the points at $\ell_C=1/3$ show a considerably lower $Nu$ 
than the DNS with a much smaller wavelength, 
a relatively good match between DNS and the experiment 
for $\ell_C=2/3$ provides some indication that at higher 
values of $\ell_C$ the saturation wavenumbers, for which $Nu \approx Nu_{fc}$ 
are smaller.}

Figure \ref{figure9b} shows two different temperatures, namely, the average
temperature just below the insulating region $\theta_{in}$ and the average
bulk temperature $\theta_{bu}$ measured at mid-height of the system.  For
$\ell_C=0.85$, $\theta_{in}$ is very close to zero, i.e. the top wall
temperature.  This is consistent with $\Nus$ nearly having the value of the
conducting case (figure \ref{figure9a}).  Also the bulk temperature is very
close to the fully conducting case $\theta=0.5$.  When $\ell_C$ is decreased,
making the top plate less and less conducting, $\theta_{in}$ increases
gradually, reaching $\theta_{in}=0.5$ when $\ell_C=0.25$.  This equals the
bulk temperature in a fully conducting system.  However, when $\ell_C$ is
decreased, also the bulk temperature gradually increases and nearly reaches
0.6 for $\ell_C=0.25$.  This rise in the bulk temperature is much slower than
the rise in the temperature above the insulating area, meaning that the
gradient between insulating regions at the plate and the bulk decreases and
thus does the heat transfer, see figure \ref{figure9a}.  Even though these
simulations were conducted using only $f=9$, one expects similar trends to
apply for other values of $f$.  From the previous simulations we found that
when increasing $f$, $\Nus/\Nus_{fc}$ will rise and $\Theta_{ins}$ will
decrease.  The same response can be achieved by increasing $\ell_C$ as we
increase the conducting area and approach the case of the fully conducting
system.

%
%%%%%%%%%% Varying the insulating fraction
\subsection{Mixed insulating and conducting patterns in two
dimensions}\label{mixedTwoDimensions} Until now, all patterns that have been
applied to the top and bottom boundary wall were one-dimensional stripe-like
patterns.  We only varied the width of the patches and the ratio between the
insulating and conducting fractions.  In this subsection we add an additional
spatial dependence to these patterns to make them checkerboard-like:
\begin{equation}
\begin{aligned}
\theta(\hat{x},\hat{y},\hat{z}=1) &= 0 &&
\hat{x} \in \left[ i L_{px},L_{px1}+i L_{px} \right], \hat{y} \in \left[ j
L_{py},L_{py1}+j L_{px} \right], i,j \in \mathbb{Z} \\ \partial_z
\theta(\hat{x},\hat{y},\hat{z}=1) &= 0 && \hat{x} \notin \left[ i
L_{px},L_{px1}+i L_{px} \right], \hat{y} \notin \left[ j L_{py},L_{py1}+j
L_{px} \right], i,j \in \mathbb{Z} \\ \theta(\hat{x},\hat{y},\hat{z}=0) &= 1
                                                                        &&
\forall \hat{x},\hat{y}.  \end{aligned} \label{eq:boundaryConditions2d}
\end{equation}

\begin{figure}
\centering
\includegraphics{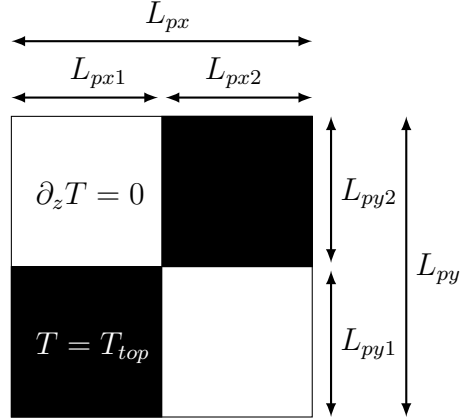}
\caption{%
Top view of the top plate boundary conditions with applied checkerboard
pattern ($f=1$).  $L_{px}$ and $L_{py}$ are the horizontal and vertical
dimensions of a set of patches.  The set itself is divided in two
insulating (white) and two conducting (black) areas, all with equal
dimensions: $L_{px1} = L_{py1} = L_{px2} = L_{py2}$.}
\label{figure12}
\end{figure}

A schematic of a set of four patches, two insulating and two conducting, is
shown in Figure \ref{figure12}.  The dimensions of both types of patches were
kept equal, i.e. $L_{px1} = L_{py1} = L_{px2} = L_{py2}$, meaning
$\ell_c=1/2$.

\begin{figure}
\centering%
\includegraphics[width=0.46\textwidth]{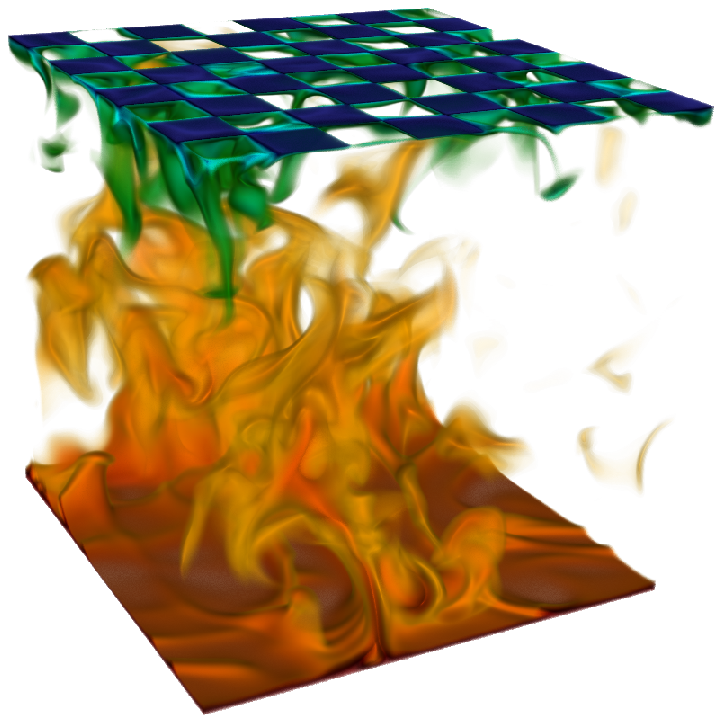}
\qquad
\includegraphics[width=0.46\textwidth]{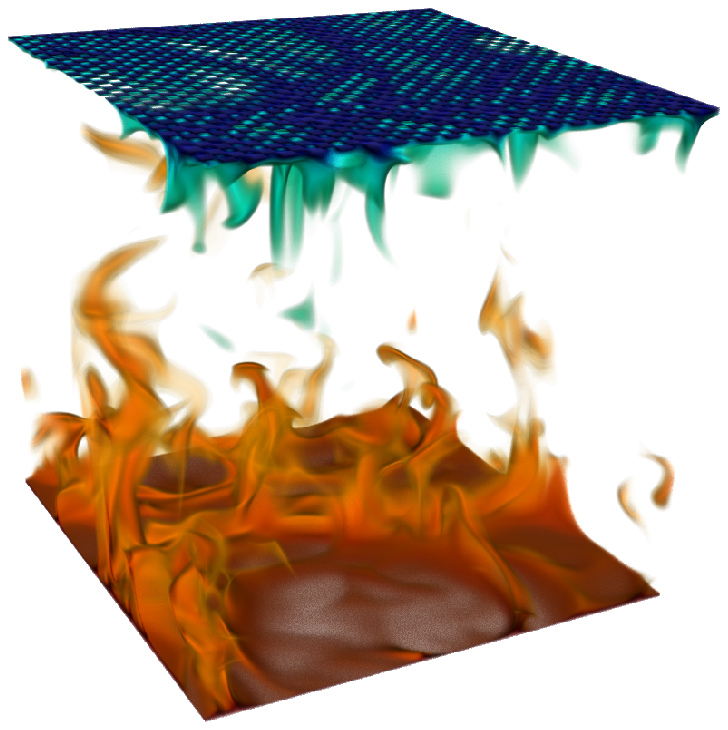}
\caption{%
Two 3D visualizations of the instantaneous temperature field with
different pattern frequencies at the top boundary. Hot fluid is shown in
red while the cool fluid has a blue color. For both visualizations,
$\Ray=10^8$ and $\Pran=1$. The left visualization shows $f=4$, which
results in 16 sets of patches containing each two insulating patches and
two conducting patches. The right visualization has $f=20$ which results
in 400 sets of patches. Hot plumes rise from the bottom plate while cold
plumes are ejected from the top boundary.}
\label{figure13}
\end{figure}

As we took both horizontal dimensions to be equal, we define a single
frequency $f$ in both directions.  The plate is divided in $f$ sets of patches
which have dimensions $L_{px}=L_{py}=L_p$.  When $f=1$ the complete boundary
consists of a single set of patches.  By increasing $f$ to 2, the plate will
consist of four sets of four patches each.  To give a further impression how
the temperature fields resulting from these boundary conditions look like, two
visualizations of instantaneous temperature fields for $f=4$ and $f=20$ are
shown in Figure \ref{figure13}.  The visualizations show respectively 16 and
400 sets of patches, which each consists of two insulating and two conducting
areas. 

\begin{figure}
\centering
\subfloat{%
\includegraphics{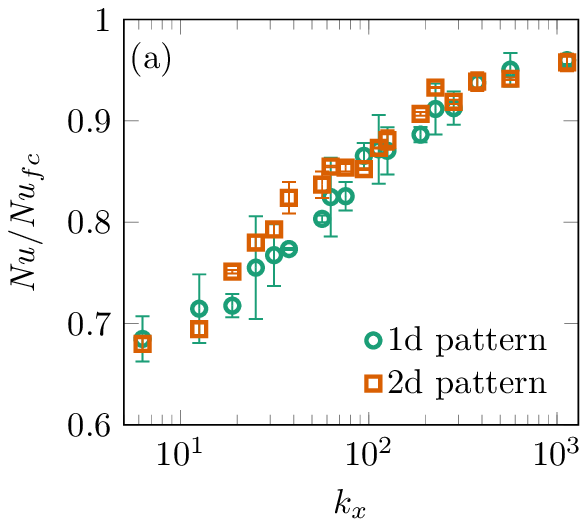}
\label{figure14a}
}
\subfloat{%
\includegraphics{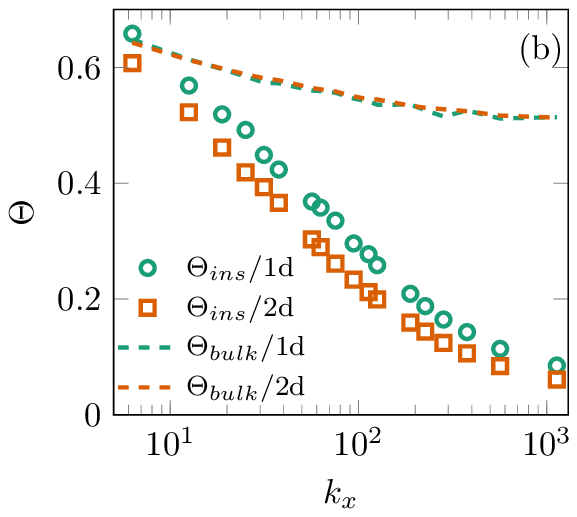}%
\label{figure14b}%
}
\caption{%
(a) Normalized Nusselt number $\Nus/\Nus_{fc}$ for 1D and 2D patterning
versus $k_x$ for $\Ray=10^8$ and
$\Pran=1$.  The error bars show the statistical convergence error.  (b)
Average temperature $\theta_{in}$ just below the insulating area and
average bulk temperature $\theta_{bu}$, both as function of $k_x$ for the
1D and 2D patternings.  The $\Ray$ and $\Pran$ are the same as for (a).  }
\label{figure14}%
\end{figure}%

Figure \ref{figure14a} shows the normalized Nusselt number $\Nus/\Nus_{fc}$
for the 1D- (stripe) and 2D (checkerboard) patterns as function of $k_x$.  For
these cases, $\Ray=10^8$, $\Pran=1$ and the error bars indicate the
statistical convergence error.  On first sight, the heat transfer is not that
different when applying one-dimensional or two-dimensional patterning.  At the
lowest division, where we have divided the system into two or four areas for
the 1D and 2D case respectively, the difference is negligible.  Also at the
highest value of $k_x$, the difference is inside statistical errors and seems
insignificant.

On first glance, the 1D stripe pattern and the 2D checkerboard pattern may
seem idealized representations. However, all other stripe and rectangular
checkerboard patterns fall within these extreme cases, as the stripe pattern
has $L_{px}/L_{py} \to \infty$ (or conversely $\to 0$), while for the
checkerboard pattern $L_{px}/L_{py}=1$.  Our work shows that, for the
relatively small wavelengths considered here, the shape of these patterns do
not have a significant influence on the flow dynamics. While we do not expect
$\ell_C$ to considerably affect this statement, larger wavelength patterns
could show some dependence on their shape. In the the experiments of
\cite{wang17}, the patches have wavelengths comparable to the system height
and their distribution considerably affects the flow structure. Here, we
consider wavelengths that are much smaller, affecting the flow only below the
thermal boundary layer thickness, and reducing the Nusselt number.  The
largest wavelength considered in this work is still smaller than the smallest
wavelength in the rectangular tank of \cite{wang17}.  There appear to be two
clear regimes: the large patch regime, which show a clear influence on the
flow dynamics and whose effect is likely to be shape-dependent, and the small
patch regime, which only affect the boundary layers and thus the heat
transfer. While there must exist a cross-over regime between both, it appears
to be just outside the wavelengths we are considering.

When we look at $\theta_{in}$, the temperature just below the insulating
region in Figure \ref{figure14b}, we see a similar trend.  The temperature
below the insulating region is on average always lower for the 2D pattern.
This is even the case for the higher and lower limit of $k_x$.  The average
temperature just below the conducing area is very close to zero, regardless of
which $k_x$.  Therefore, the distance of the colder regions is shorter for the
2D- as for the 1D-pattern.  This also explains that $\theta_{in}$ for the 2D
case is slightly lower when compared to the 1D case.  The bulk temperature for
both the 1D and 2D cases are almost identical which means that the change in
pattern has no effect on this quantity.  From these results we can conclude
that the impact of the two different patterns is very similar, and the
quantitative differences between stripe and checkerboard patterns are at most
small.

\section{Summary and conclusions}\label{Conclusion}

A series of DNS of turbulent Rayleigh-B\'enard convection using mixed
conducting- and insulating boundary conditions were conducted.  First, we
applied a stripe-like pattern on the top boundary and varied the amount of
stripes while keeping conduction-insulation ratio constant at $\ell_C=1/2$.
When the top plate is divided in half, $\Nus$ has a value of approximately
two-thirds of the fully conducting case.  By increasing the frequency of the
pattern, the $\Nus$ also increases, with a maximum value very close to as if
it were fully conducting.  With only half the effective conducting area on the
top plate, when applying a dense pattern, the effect of the insulating patches
almost completely vanishes.  An increase in $\Ray$ results only in a marginal
decrease in $\Nus/\Nus_{fc}$ for the largest $f$.  This shift towards the
fully conducting efficiency as seen with the Nusselt number when increasing
the pattern frequency is also visible in $\theta_{bulk}$ and $\theta_{ins}$.

Using a two-dimensional Fourier transform, calculated from a horizontal slice
of the instantaneous temperature it is possible to identify the imprint of the
boundary conditions inside the flow.  By comparing different spectra, each
calculated from a horizontal slice slightly further away from the top boundary
wall, the penetration depth of the boundary conditions inside the flow was
investigated.  The imprint of the striped pattern slowly fades away when
moving from the top wall towards the border of the thermal boundary layer.
Outside of the thermal boundary layer the imprint has completely vanished,
even for the extreme case $f=1$.  The thermal boundary layer masks the actual
boundary, including all insulating imperfections and presents a new effective
boundary to the bulk flow.  In the thermal boundary layer, the heat is
conducted to the conducting areas.  This transport is more efficient when the
pattern frequency is large.  A lower Rayleigh number increases the thickness
of the boundary layer and thereby, also increases the effectiveness of the
heat transport.

Extending the pattern to both, the top and bottom boundary wall, resulted in
similar behavior for the heat transfer and the average temperature below the
insulating area.  The primary difference is for the lowest pattern frequency
where we practically have only half a RB cell and we find a Nusselt number
with half the value of the fully conducting case.  Adding an additional
dimension to the pattern, creating a checkerboard-like pattern, also did not
change the behavior significantly.

Our results demonstrate that small and even large imperfections in the
temperature boundary conditions are barely felt in the system dynamics in
terms of global heat transfer and local temperature measurements.  Only in
extreme cases as a half-and-half conducting and adiabatic plate was the effect
significant. The effect of imperfect temperature boundary conditions of fully
turbulent RB is weaker than the effects of velocity boundary conditions in
two-dimensional RB \citep{poe14}, or the effect of rough elements near the
boundaries \citep{tis11}. It is not yet clear if these boundary imperfections
lead to significant changes in the dynamics of the bulk flow and this remains
an open question for future works. Going beyond the scope of
 the present paper, we mention that the simulations 
by \cite{cooper13} and the  experiments by \cite{wang17} show that 
with even larger adiabatic patches, changes in the flow topology can 
happen due to the arrangement of patches. The patches can also be varied in
time, which is a way to control the bulk temperature or to fine tune the heat
transfer, which is relevant for many industrial applications.
In another work \cite{whitehead15}
showed that the combination of shear and mixed boundary conditions could
also play a critical role in the system dynamics. Understanding the deeper
reasons for this behavior may lead to better models for natural convection for
geo- and astro-physically relevant flows.

\emph{Acknowledgments}: We thank L. Biferale for his role in motivating this
manuscript. We also thank V. Spandan, Y. Yang and X. Zhu for fruitful
discussions. D.B. and E.P. were funded by FOM grants, and R.O. was funded by
an ERC Advanced Grant. Computing time at the Dutch supercomputer Cartesius
comes from a NWO grant.

\bibliographystyle{jfm}
\bibliography{JFM_MixedBCs}

%%%%%%%%%%%%%%%%%%%%%%%%%%%%%
%      Appendix Tables      %
%%%%%%%%%%%%%%%%%%%%%%%%%%%%%

\newpage

\appendix \section{Numerical details}\label{appendixA}

\begin{center}
\begin{tabular}{cccc}
\toprule
\multicolumn{4}{c}{$\text{Ra}=10^7$,$\text{Pr}=1$, and $\ell_c=1/2$} \\
	$f$ & $\text{Nu}$ & $\Theta_{bulk}$ & $\Theta^{top}_{ins}$ \\
\midrule
	1 & $11.82\pm0.18$ & 0.67 & 0.69 \\
	2 & $12.44\pm0.31$ & 0.62 & 0.55 \\
	3 & $12.75\pm0.27$ & 0.61 & 0.48 \\
	4 & $13.76\pm0.01$ & 0.60 & 0.43 \\
	5 & $13.86\pm0.02$ & 0.59 & 0.40 \\\\
	9 & $14.77\pm0.07$ & 0.57 & 0.29 \\
	10 & $14.56\pm0.03$ & 0.57 & 0.28 \\
	12 & $14.97\pm0.15$ & 0.56 & 0.25 \\
	15 & $14.81\pm0.12$ & 0.55 & 0.22 \\
	18 & $15.54\pm0.15$ & 0.55 & 0.19 \\
	20 & $15.68\pm0.07$ & 0.55 & 0.18 \\
	30 & $15.98\pm0.15$ & 0.53 & 0.14 \\
	36 & $16.02\pm0.15$ & 0.53 & 0.12 \\
	45 & $16.07\pm0.14$ & 0.52 & 0.10 \\
	60 & $16.60\pm0.05$ & 0.52 & 0.08 \\
	90 & $16.58\pm0.15$ & 0.52 & 0.06 \\
	180 & $16.95\pm0.18$ & 0.51 & 0.05 \\
	fc & $16.90\pm0.07$ & 0.50 & -- \\
\bottomrule
\end{tabular}%
\begin{tabular}{cccc}
\toprule
\multicolumn{4}{c}{$\text{Ra}=10^8$,$\text{Pr}=1$, and $\ell_c=1/2$} \\
	$f$ & $\text{Nu}$ & $\Theta_{bulk}$ & $\Theta^{top}_{ins}$ \\
\midrule
	1 & $22.08\pm0.72$ & 0.65 & 0.66 \\
	2 & $23.04\pm1.09$ & 0.61 & 0.57 \\
	3 & $23.14\pm0.37$ & 0.60 & 0.52 \\
	4 & $24.34\pm1.64$ & 0.58 & 0.49 \\
	5 & $24.74\pm0.98$ & 0.57 & 0.45 \\
	6 & $24.94\pm0.03$ & 0.57 & 0.42 \\
	9 & $25.90\pm0.10$ & 0.56 & 0.37 \\
	10 & $26.59\pm1.25$ & 0.56 & 0.36 \\
	12 & $26.62\pm0.46$ & 0.56 & 0.34 \\
	15 & $27.90\pm0.41$ & 0.55 & 0.30 \\
	18 & $28.11\pm1.09$ & 0.54 & 0.28 \\
	20 & $28.06\pm0.75$ & 0.54 & 0.26 \\
	30 & $28.58\pm0.24$ & 0.54 & 0.21 \\
	36 & $29.39\pm0.81$ & 0.53 & 0.19 \\
	45 & $29.42\pm0.53$ & 0.52 & 0.16 \\
	60 & $30.23\pm0.20$ & 0.53 & 0.14 \\
	90 & $30.64\pm0.53$ & 0.51 & 0.11 \\
	180 & $30.95\pm0.12$ & 0.51 & 0.09 \\
	fc & $32.87\pm0.33$ & 0.50 & -- \\
\bottomrule
\end{tabular}
\begin{tabular}{cccc}
\toprule
\multicolumn{4}{c}{$\text{Ra}=10^8$,$\text{Pr}=10$, and $\ell_c=1/2$} \\
	$f$ & $\text{Nu}$ & $\Theta_{bulk}$ & $\Theta^{top}_{ins}$ \\
\midrule
	1 & $22.92\pm0.18$ & 0.65 & 0.67 \\
	2 & $24.04\pm0.23$ & 0.62 & 0.58 \\
    3 & $23.84\pm0.11$ & 0.60 & 0.53 \\
    4 & $24.86\pm0.03$ & 0.59 & 0.50 \\
	5 & $26.55\pm0.88$ & 0.58 & 0.48 \\
    6 & $25.32\pm0.80$ & 0.58 & 0.46 \\
	9 & $26.94\pm0.18$ & 0.57 & 0.37 \\
	10 & $26.71\pm0.62$ & 0.57 & 0.37 \\
    12 & $27.58\pm0.40$ & 0.56 & 0.34 \\
	15 & $27.79\pm0.15$ & 0.55 & 0.31 \\
	18 & $27.66\pm0.28$ & 0.55 & 0.28 \\
    20 & $28.80\pm0.19$ & 0.55 & 0.26 \\
	30 & $29.06\pm0.09$ & 0.54 & 0.22 \\
	36 & $28.96\pm1.01$ & 0.54 & 0.2 \\
	45 & $29.33\pm0.17$ & 0.53 & 0.17 \\
	60 & $30.27\pm0.38$ & 0.53 & 0.14 \\
    90 & $29.98\pm1.10$ & 0.52 & 0.11 \\
    180 & $30.60\pm0.54$ & 0.52 & 0.08 \\
    fc & $32.63\pm0.13$ & 0.50 & -- \\
\bottomrule
\end{tabular}%
\begin{tabular}{cccc}
\toprule
\multicolumn{4}{c}{$\text{Ra}=10^9$,$\text{Pr}=1$, and $\ell_c=1/2$} \\
	$f$ & $\text{Nu}$ & $\Theta_{bulk}$ & $\Theta^{top}_{ins}$ \\
\midrule
	1 & $43.87\pm0.22$ & 0.65 & 0.65 \\
	2 & $45.05\pm0.27$ & 0.62 & 0.60 \\
	3 & $45.94\pm0.19$ & 0.61 & 0.56 \\
	4 & $47.14\pm0.09$ & 0.60 & 0.53 \\
	5 & $47.66\pm0.28$ & 0.59 & 0.50 \\\\
	9 & $49.15\pm0.28$ & 0.58 & 0.44 \\
	10 & $50.24\pm1.41$ & 0.58 & 0.43 \\
	12 & $51.46\pm0.73$ & 0.57 & 0.41 \\
	15 & $51.48\pm0.05$ & 0.57 & 0.38 \\
	18 & $52.87\pm0.32$ & 0.56 & 0.36 \\
	20 & $53.54\pm0.97$ & 0.56 & 0.35 \\
	30 & $54.84\pm0.37$ & 0.55 & 0.30 \\
	36 & $55.7\pm0.03$ & 0.55 & 0.28 \\
	45 & $55.71\pm0.20$ & 0.54 & 0.26 \\
	60 & $56.52\pm0.92$ & 0.54 & 0.23 \\
	90 & $58.36\pm0.38$ & 0.53 & 0.20 \\
	180 & $59.42\pm0.68$ & 0.52 & 0.16 \\
	fc & $64.66\pm0.33$ & 0.50 & -- \\
\bottomrule
\end{tabular}%
\captionof{table}{Data from all numerical simulations for the single-sided
mixed boundary conditions and the fully conducting case, used in figures 4a,
5a, 5b, and 8b. The ‘fc’ in the f column indicates the fully conducting case,
e.g. without the striped pattern.}
\end{center}

\begin{center}
\begin{tabular}{ccccc}
    \toprule
    \multicolumn{5}{c}{$\text{Ra}=10^8$,$\text{Pr}=1$, and $\ell_c=1/2$} \\
	$f$ & $\text{Nu}$ & $\Theta_{bulk}$ & $\Theta^{top}_{ins}$ &
    $\Theta^{bot}_{ins}$ \\
    \midrule
	1 & $16.96\pm0.11$ & 0.52 & 0.55 & 0.53 \\
	2 & $17.67\pm0.03$ & 0.52 & 0.56 & 0.49 \\
	4 & $19.10\pm0.33$ & 0.55 & 0.64 & 0.45 \\
	10 & $21.98\pm0.17$ & 0.53 & 0.71 & 0.34 \\
	15 & $23.60\pm0.39$ & 0.53 & 0.74 & 0.28 \\
	20 & $24.77\pm0.26$ & 0.52 & 0.78 & 0.25 \\
	36 & $26.93\pm0.42$ & 0.51 & 0.83 & 0.18 \\
	60 & $27.76\pm0.98$ & 0.50 & 0.87 & 0.14 \\
	90 & $29.03\pm0.33$ & 0.50 & 0.89 & 0.11 \\
	180 & $30.10\pm0.97$ & 0.51 & 0.92 & 0.08 \\
    \bottomrule
\end{tabular}
\captionof{table}{Data from all numerical simulations for the double sided
mixed boundary conditions, used in figure 8a and 8b.}
\end{center}

\begin{center}
\begin{tabular}{cccc}
\toprule
\multicolumn{4}{c}{$\text{Ra}=10^8$,$\text{Pr}=1$, and $\ell_c=1/2$} \\
	$f$ & $\text{Nu}$ & $\Theta_{bulk}$ & $\Theta^{top}_{ins}$ \\
\midrule
	1 & $22.35\pm0.26$ & 0.64 & 0.61 \\
	2 & $22.82\pm0.21$ & 0.61 & 0.52 \\
	3 & $24.69\pm0.06$ & 0.60 & 0.46 \\
	4 & $25.64\pm0.19$ & 0.59 & 0.42 \\
	5 & $26.06\pm0.20$ & 0.58 & 0.39 \\
	6 & $27.09\pm0.51$ & 0.58 & 0.37 \\
	9 & $27.51\pm0.43$ & 0.56 & 0.30 \\
	10 & $28.10\pm0.02$ & 0.56 & 0.29 \\
	12 & $28.06\pm0.12$ & 0.56 & 0.26 \\
	15 & $28.02\pm0.03$ & 0.55 & 0.23 \\
	18 & $28.71\pm0.02$ & 0.55 & 0.21 \\
	20 & $28.96\pm0.32$ & 0.54 & 0.20 \\
	30 & $29.81\pm0.06$ & 0.53 & 0.16 \\
	36 & $30.66\pm0.01$ & 0.53 & 0.14 \\
	45 & $30.20\pm0.07$ & 0.53 & 0.12 \\
	60 & $30.86\pm0.31$ & 0.52 & 0.11 \\
	90 & $30.95\pm0.04$ & 0.52 & 0.08 \\
	180 & $31.48\pm0.28$ & 0.51 & 0.06 \\ 
\bottomrule
\end{tabular}
\captionof{table}{Data from all numerical simulations of the checkerboard
pattern used in figure 12a and 12b.}
\end{center}

\begin{center}
\begin{tabular}{ccccccc}
\toprule
\multicolumn{7}{c}{$\text{Pr}=1$, and $\ell_c=1/2$} \\
	$f$ & $\text{Ra}$ & $\text{Nu}$ & &
	$f$ & $\text{Ra}$ & $\text{Nu}$ \\
\midrule
    1 & $\num{1e7}$ & 11.82 & & 90 & $\num{1e9}$ & 58.36 \\
    1 & $\num{1e8}$ & 22.08 & & 90 & $\num{3e9}$ & 81.37 \\
    1 & $\num{3e8}$ & 30.18 & & 180 & $\num{1e7}$ & 16.95 \\
    1 & $\num{1e9}$ & 43.87 & & 180 & $\num{1e8}$ & 30.95 \\
    1 & $\num{3e9}$ & 63.26 & & 180 & $\num{3e8}$ & 42.43 \\
	10 & $\num{1e7}$ & 14.56 & & 180 & $\num{1e9}$ & 59.42 \\
	10 & $\num{1e8}$ & 26.59 & & 180 & $\num{3e9}$ & 82.53 \\
	10 & $\num{3e8}$ & 35.58 & & fc & $\num{1e7}$ & 16.99 \\
	10 & $\num{1e9}$ & 50.24 & & fc & $\num{1e8}$ & 32.87 \\
	10 & $\num{3e9}$ & 70.73 & & fc & $\num{3e8}$ & 44.71 \\
	90 & $\num{1e7}$ & 16.58 & & fc & $\num{1e9}$ &  63.95 \\
	90 & $\num{1e8}$ & 30.64 & & fc & $\num{3e9}$ &  92.15 \\
	90 & $\num{3e8}$ & 41.34 & & & & \\
\bottomrule
\end{tabular}
\captionof{table}{Nusselt and Rayleigh numbers for various $f$ and the fully
conducting case, used in figure 3a and 3b}
\end{center}

\end{document}